\documentclass[pr,twocolumn,preprintnumbers,nofootinbib,showpacs,amsmath,amssymb,superscriptaddress]{revtex4-1}
\usepackage{amsfonts}
\usepackage{mathrsfs}
\usepackage{amssymb}
\usepackage{amsmath}
\usepackage{color}
\usepackage[
      colorlinks=true,
      linkcolor=blue,
      urlcolor=blue,
      filecolor=black,
      citecolor=blue,
            ]{hyperref}

\usepackage{amsfonts}
\usepackage{graphicx}
\usepackage{epstopdf}
\usepackage{bigstrut,bigdelim,multirow}

\begin{document}
\title{Holographic antiferromganetic quantum criticality and AdS$_2$ scaling limit}
\author{Rong-Gen Cai}
\email{cairg@itp.ac.cn}

\author{Run-Qiu Yang}
\email{aqiu@itp.ac.cn}
\affiliation{State Key Laboratory of Theoretical Physics,Institute of Theoretical Physics,\\
 Chinese Academy of Sciences,Beijing 100190, China.}

\author{F.V. Kusmartsev}
\email{F.Kusmartsev@lboro.ac.uk}
\affiliation{Department of Physics, Loughborough University, Loughborough, Leicestershire, LE11 3TU, United Kingdom}


\begin{abstract}
A holographic description on antiferromagnetic quantum phase transition (QPT) induced by magnetic field and the criticality in the vicinity of quantum critical point (QCP) have been investigated numerically recently. In this paper, we show that the properties of QPT in this holographic model are governed by a CFT dual to the emergent AdS$_2$ in the IR region, which confirms that the dual boundary theory is a strong coupling theory with dynamic exponent $z=2$ and logarithmic corrections appear. We also compare them with the results from Hertz model by solving RG equation at its upper critical dimension and with some experimental data from pyrochlores Er$_{2-2x}$Y$_{2x}$Ti$_2$O$_7$ and   BiCoPO$_5$.
\end{abstract}
 \maketitle

\noindent

\section{Introduction}
Quantum phase transitions (QPTs) happen at zero temperature. Such a phase transition and the critical behavior in the vicinity of the corresponding quantum critical points (QCPs)  have attracted a great deal of interest both in theory and  experiment sides recently~\cite{S. Sachdev,Hertz.76,Sachdev_book,Natphys.08}. In phase diagram, A QCP separates an ordered phase from a disordered phase at zero temperature. In contrast to classical phase transition at $T>0$ where thermal fluctuations play important role, QPTs are induced  by  quantum fluctuations associated with the Heisenberg uncertainty and driven by a control parameter in Hamiltonian rather than temperature. Usually, the control  parameter could be composition, magnetic field, or pressure, etc.. In condensed matter physics, such a quantum criticality is considered to be an important mechanism for some of the most interesting phenomena, especially in itinerant electronic systems \cite{Gegenwart.08,Lohneysen.07} and other phenomenons involving strongly correlated many-body systems~\cite{Si.01,Senthil.04}. However, the complete theoretical descriptions valid at all the energy (or temperature) region are still lacking.

\begin{figure}[h]
\begin{center}
\includegraphics[width=0.4\textwidth]{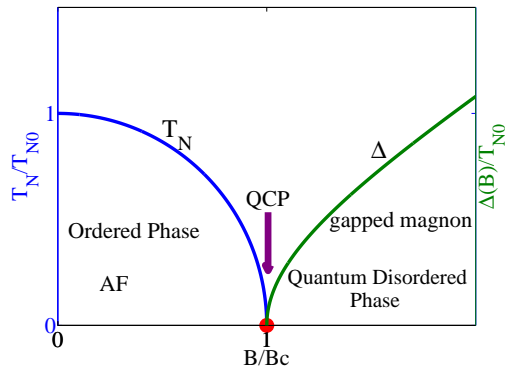}
\caption{Schematic  phase diagram for a QPT, showing the N\'{e}el temperature and spin gap  as  functions of magnetic field~\cite{Ch.08}. For a magnetic QPT, the characteristic energy scales in the QD (quantum disordered) and AF phases are, respectively, the spin gap $\Delta$ and N\'{e}el temperature $T_N$, both vanish at the QCP.}
\label{figTDg}
\end{center}
\end{figure}

One of the most interesting and intensively  discussed QPTs is the ordered-disordered QPT in antiferromagnetic materials induced by magnetic field, such as Cu$_2$(C$_5$H$_12$N$_2$)$_2$Cl$_4$~\cite{G.Chaboussant}, KCuCl$_3$~\cite{W.Shiramura}, TlCuCl$_3$~\cite{Ch.03}, BiCoPO$_5$~\cite{E.Mathews} and Er$_{2-2x}$Y$_{2x}$Ti$_2$O$_7$~\cite{Niven}. A  schematic phase diagram in the vicinity of a QPT as a function of magnetic field is shown in Fig~\ref{figTDg}, where we plot the N\'{e}el temperature $T_N(B)$ when $B<B_c$ and the spin gap $\Delta(B)$ at  zero temperature when $B>B_c$. In the antiferromagnetic (AF) phase (ordered phase), the uniform spontaneously staggered magnetization is accompanied by a long-ranged spin ordering, which persists up to a finite transition temperature $T_N$. When magnetic field increases, the  transition temperature $T_N$ is suppressed. When the magnetic field arrives at its critical value, $B=B_c$, the transition temperature is suppressed to zero. When $B>B_c$ and $T=0$, the quantum disordered (QD) phase has gapped magnetic excitations.

Another interesting feature in QPTs involving strong coupling is so called ``hyperscaling violation" when the effective dimension is larger than or equal to $d_{c}$, the upper critical dimension of corresponding field theory. In general for the case of dimension less than $d_{c}$, in the vicinity of QCP, there are the hyperscaling relations for free energy density $F(b,T)$, N\'{e}el temperature $T_N$, correlation length $\xi$ and characteristic energy $\Delta$ with respect to the tuning parameter $b=B/B_c-1$ as follows,
\begin{subequations}\label{hyperscal}
\begin{align}
&F_c(b,T)=\lambda^{-(d+z)}F_c(b\lambda^{1/\nu},T\lambda^z),~~b\rightarrow0\label{hyp1}\\
&T_N\propto(-b)^\psi,~~b\rightarrow0^-,\label{hyp2}\\
&\xi\propto b^{-\nu}, ~\Delta\propto b^{z\nu} ~~b\rightarrow0^+,\label{hyp3}
\end{align}
\end{subequations}
where $d$ is the spatial dimension of the system, $z$ is called dynamic exponent, $\psi$ and $\nu$ are two positive critical exponents depending on model. $F_c(b,T)$ is the critical contribution to the free energy density defined by $F_c(b,T)=F(b,T)-F_{\text{reg}}(b,T)$ and $\lambda$ is an arbitrary scale factor. The effective dimension of the system at QPT then is $d_{\text{eff}}=d+z$. The scaling relations~\eqref{hyp3} are always valid. But the naive scaling relation~\eqref{hyp1} is valid only when the effective dimension is less than  the upper critical dimension, i.e., $d_{\text{eff}}<d_c$~\cite{Lohneysen.07}. In particular,  when $d=z=2$, the hyperscaling relations~\eqref{hyp1} and~\eqref{hyp2} are no longer valid. In that case, the quantum critical theory is not in a weak-coupling region in general. So for this case, a description of a strongly coupled effective field theory is called for near the QPTs~\cite{Lohneysen.07}.

In order to study and characterize strongly coupled quantum critical systems, some new methods are needed.  The gauge/gravity duality or AdS/CFT correspondence provides us with  a promising approach~\cite{Maldacena:1997re,Gubser:1998bc,Witten:1998qj,Witten:1998qj2}. This duality relates a weak coupling gravitational theory in  a $(d+1)$-dimensional asymptotically anti de-Sitter (AdS) space-time to a $d$-dimensional strong coupling conformal field theory (CFT) in the AdS boundary. In recent years, this duality has been extensively applied into condensed matter systems. Some remarkable  progresses have been made in this direction, including holographic superfluids (superconductors)~\cite{Gubser,Hartnoll:2008vx} (for a recent review, see \cite{Cai:2015cya}), (non-)Fermi liquids~\cite{Lee:2008xf,Liu:2009dm,Cubrovic:2009ye} and so on. The models in the AdS/CFT frame for ferromagnetism/paramagnetism and anti-ferromagnetism/paramagnetism phase transitions have also been proposed in~\cite{Cai:2014oca,Cai:2014jta}. Especially in Ref.~\cite{Cai:2014jta}, we found that the antiferroamgnetic transition temperature $T_N$ is indeed suppressed by an external magnetic field and tends to zero when the magnetic field approaches to a critical value $B_c$. In this way we realized a holographic description of antiferromagnetic quantum phase transition induced by an external magnetic field.  Since the model involves  rank-2 fields, the problems about ghost and causal violation may appear, which have not been discussed. Very recently, a modified model based on the previous works has been proposed in Ref.~\cite{Cai:2015bsa}, which  is shown to be ghost free and causal violation is absent,  but keeps all the significant results in  the previous works qualitatively. For this modified model, a numerical investigation about antiferromagnetic quantum phase transition induced by an external magnetic field was first presented in Ref.~\cite{Cai:2015a}.

In this paper, we will further elaborate on the magnetic induced QPT and study its critical behavior in some details.  In  section~\ref{AFM}, we will first review some basic properties of holographic antiferromagnetic model by combining the ideas in Refs.~\cite{Cai:2014jta,Cai:2015bsa} and study the behavior of N\'{e}el temperature $T_N$ with respect to the external magnetic field, which shows that there is a critical magnetic field where quantum criticality appears and can be fitted well by adding a logarithmic correction to the usual power law form. In  section~\ref{AFM2}, we will show there is an emergent AdS$_2$ scaling limit which governs the low frequency behaviors in the vicinity of QCP. By this emergent AdS$_2$ geometry, we confirm the numerical results in section~\ref{AFM} and Ref.~\cite{Cai:2015a}. In  section~\ref{AFM3}, we give a  dual field computation from perturbation RG equation inspired by holographic results, which gives a confirmation about our holographic results. We will also discuss what our model can tell us about some real materials in this section.

\section{antiferromagnetic model and critical temperature}
\label{AFM}
Antiferromagnetic phase in materials is a magnetic ordered phase without net magnetization. For the simplest case, there are two different sub-lattices which have two spontaneous magnetic moments with the same magnitude but opposite direction. The order parameter then is the staggered magnetization, i.e., the difference between two different magnetic moments in two sub-lattices. The same as in the ferromagnetic phase, the time reversal transformation is also broken spontaneously in the antiferromagnetic phase.

\subsection{Holographic antiferromagnetic model}
In  Ref.~\cite{Cai:2014oca}, we introduced an antisymmetric tensor field coupling with U(1) strength field to describe the phase which has one uniform spontaneous magnetization. In order to describe a spontaneous magnetic ordered phase which has a staggered magnetization, we introduce two antisymmetric tenser fields. Following Ref.~\cite{Cai:2014jta} and Ref.~\cite{Cai:2015bsa}, the action we will consider here takes the form
\begin{equation}\label{action1}
S=\frac1{2\kappa^2}\int d^4x\sqrt{-g}[R+\frac{6}{L^2}-F^{\mu\nu}F_{\mu\nu}+\lambda^2(L_1+L_2+L_{12})],
\end{equation}
where
\begin{equation}\label{action2}
\begin{split}
&L_{12}=\frac k2M^{(1)\mu\nu}M^{(2)}_{\mu\nu},\\
&L_{(a)}=\frac1{12}(dM^{(a)})^{\mu\nu\tau}(dM^{(a)})_{\mu\nu\tau}+\frac{m^2}4 M^{(a)\mu\nu}M^{(a)}_{\mu\nu}\\
&~~~~~~~~~+\frac12M^{(a)\mu\nu}F_{\mu\nu}+J V(M^{(a)}_{\mu\nu}),\\
&~V(M^{(a)}_{\mu\nu})=({^*M^{(a)}}_{\mu\nu}M^{(a)\mu\nu})^2, ~a=1,2.
\end{split}
\end{equation}
$L$ is the radius of AdS space, $2\kappa^2=16\pi G$ with $G$  the Newton constant. $^*$ is the Hodge star dual operator. In this model, $k$, $m^2$ and $J$ are all model parameters with $J<0$. $\lambda^2$ characterizes the back reaction of the two polarization fields $M^{(a)}_{\mu\nu}$ with $a=1,2$ to the background geometry, and $L_{12}$ describes the interaction between two polarization fields. The equations of motion for polarization fields read
\begin{equation}\label{eomFM}
3\nabla^\tau\nabla_{[\tau}M^{(a)}_{\mu\nu]}-m^2M^{(a)}_{\mu\nu}-kM^{(b)}_{\mu\nu}-JV'{(a)}_{\mu\nu}-F_{\mu\nu}=0.
\end{equation}
Here $(a,b)=(1,2)$ or $(2,1)$ and $V'{(a)}_{\mu\nu}=({^*M^{(a)}}_{\tau\sigma}M^{(a)\tau\sigma}){^*M^{(a)}}_{\mu\nu}$. In the probe limit of $\lambda\rightarrow0$, we can neglect the back reaction of the two polarization fields on the background geometry. The background we will consider is a dyonic Reissner-Nordstr\"om-AdS black brane solution of
the Einstein-Maxwell theory with a negative cosmological constant, and the metric reads~\cite{Cai:1996eg}
\begin{equation}\label{geom}
\begin{split}
  ds^2=r^2(-f(r)dt^2+dx^2+dy^2)+\frac{dr^2}{r^2f(r)},\\
   f(r)=1-\frac{1+\mu^2+B^2}{r^3}+\frac{\mu^2+B^2}{r^4}.
\end{split}
\end{equation}
Here both the black brane horizon and AdS radius have been set to be unity. The temperature of the black
brane is
\begin{equation}\label{Tem1}
T=\frac1{4\pi}(3-\mu^2-B^2).
\end{equation}
For the solution (\ref{geom}), the corresponding gauge potential is $ A_\mu=\mu(1-1/r)dt+Bx dy$. Here $\mu$ is the chemical potential and $B$ can be viewed as the external magnetic field in the dual boundary field theory. The zero temperature limit then corresponds to $B^2+\mu^2=3$.

In order to describe antiferromagnetic phase transition, following  Ref.~\cite{Cai:2014jta}, we consider  $M^{(a)}_{tr}, M^{(a)}_{xy}$ ($a=1,2$) and define
\begin{equation}\label{ab1}
\begin{split}
\alpha=\frac12(M^{(1)}_{xy}+M^{(2)}_{xy}),~~\beta=\frac12(M^{(1)}_{xy}-M^{(2)}_{xy}),\\
p_1=\frac12(M^{(1)}_{tr}+M^{(2)}_{tr}),~~p_2=\frac12(M^{(1)}_{tr}-M^{(2)}_{tr}).
\end{split}
\end{equation}
Then different values of $\alpha$ and $\beta$ correspond to different magnetic phases.  The staggered magnetization $N^\dag$ and total magnetization $N$ can be defined as,
\begin{equation}\label{staggN}
\begin{split}
&N^\dag/\lambda^2=-\int_1^{\infty}\frac{\beta}{r^2}dr,\\
&N/\lambda^2=-\int_1^{\infty}\frac{\alpha}{r^2}dr.
\end{split}
\end{equation}
When external magnetic field $B=0$, the antiferromagnetic phase corresponds to the phase with nonzero staggered magnetization but zero total magnetization density.

Put the expressions~\eqref{ab1} into equations~\eqref{eomFM}, we have the equations for $\alpha$ and $\beta$ as
\begin{equation}\label{eqab}
\begin{split}
\alpha''+\frac{f'\alpha'}f-\left[\frac{4J(p_1^2+p_2^2)}{r^2f}+\frac{m^2+k}{r^2f}\right]\alpha&\\
+\frac{8Jp_1p_2\beta}{r^2f}+\frac{B}{r^2f}=0,&\\
\beta''+\frac{f'\beta'}f-\left[\frac{4J(p_1^2+p_2^2)}{r^2f}+\frac{m^2-k}{r^2f}\right]\beta&\\
+\frac{8Jp_1p_2\alpha}{r^2f}=0&
\end{split}
\end{equation}
with
\begin{equation}\label{eqabp}
\begin{split}
p_1&=\frac{r^2[(m^2-k)r^4-4J(\alpha^2+\beta^2)]\mu}{16J^2(\alpha^2-\beta^2)^2-8J(\alpha^2+\beta^2)m^2r^4+(m^4-k^2)r^8},\\
p_2&=\frac{8J\mu\alpha\beta r^2}{16J^2(\alpha^2-\beta^2)^2-8J(\alpha^2+\beta^2)m^2r^4+(m^4-k^2)r^8},
\end{split}
\end{equation}

The behavior of the solutions of Eqs.~\eqref{eqab} in the UV region (near the AdS boundary) depends on the value of $m^2+k$. When $m^2+k=0$, the asymptotic solutions will have a logarithmic term, we will not consider this case in present paper.  On the other hand, when $m^2+k\neq0$, we have the solution:
\begin{equation}\label{ab2}
\begin{split}
&\alpha_{UV}=\alpha_+^{UV}r^{(1+\delta_1)/2}+\alpha_-^{UV}r^{(1-\delta_1)/2}-\frac{B}{m^2+k},\\
&\beta_{UV}=\beta_+^{UV}r^{(1+\delta_2)/2}+\beta_-^{UV}r^{(1-\delta_2)/2},\\
&\delta_1=\sqrt{1+4k+4m^2},~~\delta_2=\sqrt{1-4k+4m^2},
\end{split}
\end{equation}
where $\alpha_{\pm}^{UV}$ and $\beta_{\pm}^{UV}$ are all finite constants. We require that the condensation happens spontaneously when $B=0$, by the inspirit of AdS/CFT, which gives boundary conditions such that
\begin{equation}\label{cond0}
    \alpha_+^{UV}=\beta_+^{UV}=0.
\end{equation}

\subsection{Conditions for antiferromagnetic phase transition when B=0}
In our model, there are three parameters $J, m^2$ and $k$. In order  that the dual boundary theory can describe antiferromagnetic system, these parameters need satisfy some conditions.

First, from the asymptotic behaviors in Eqs.~\eqref{ab2}, we see  that the BF-bound at the boundary requires that $m^2$ and $k$ satisfy $1+4m^2>4|k|$. On the other hand, because of $J<0$, in order to make $p_1$ and $p_2$ finite in the region $(r_h,\infty)$, we need following conditions,
\begin{equation}\label{cond2}
m^2>|k|\Leftrightarrow\delta_1>1,~\text{and}~\delta_2>1.
\end{equation}
If conditions~\eqref{cond2} are broken, we can see that the components $p_1, p_2$ will diverge at somewhere, which leads that there is some point where the energy-momentum tensor  associated with
these tensor fields diverges. So in this case, the bulk geometry becomes unstable and the probe limit is broken.

We can explain  the conditions~\eqref{cond0} and conditions~\eqref{cond2} in other way. By the definitions~\eqref{ab1}, we see that the terms of $\alpha_{\pm}^{UV}, \beta_{\pm}^{UV}$ and $B$ all appear in the asymptotic solutions of $M_{xy}^{(1)}$ and $M_{xy}^{(1)}$.  Since we need that the system is dominated by external magnetic field when $B\neq0$, so the magnetic term should be the leading term  near the boundary. This can be reached only when both the boundary conditions~\eqref{cond0} and conditions~\eqref{cond2} are satisfied.

Second, as pointed  out in Ref.~\cite{Cai:2014jta}, in order to get an antiferromagnetic phase transition below a critical temperature when $B=0$, we should achieve that $\beta$ can condense spontaneously,  but  $\alpha$ cannot, or  in other words, the system is stable for $\alpha$ but not  for $\beta$. In RN-AdS black brane background, this can be achieved if the AdS$_2$ BF-bound is satisfied for $\alpha$ but violated for $\beta$ near the horizon when $T=0$. As in the critical cases for instability, we can treat $\alpha$ and $\beta$ both infinitesimal, so we get a linearized equations for them,
\begin{equation}\label{eqab2}
\begin{split}
\alpha''+\frac{f'\alpha'}f-\left[\frac{4Jp_1^2}{r^2f}+\frac{m^2+k}{r^2f}\right]\alpha=0,&\\
\beta''+\frac{f'\beta'}f-\left[\frac{4Jp_1^2}{r^2f}+\frac{m^2-k}{r^2f}\right]\beta=0&
\end{split}
\end{equation}
with $p_1=\mu/[(m^2+k)r^2]$. Then zero temperature corresponds to $\mu=\sqrt{3}$. Let $r_*=1/(r-1)$, when $r_*\rightarrow\infty$, we have following asymptotic solutions for $\alpha$ and $\beta$,
\begin{equation}\label{ab2c}
\begin{split}
&\alpha_{IR}=\alpha_+^{IR}r_*^{-(1+\delta_1^*)/2}+\alpha_-^{IR}r_*^{-(1-\delta_1^*)/2},\\
&\beta_{IR}=\beta_+^{IR}r_*^{-(1+\delta_2^*)/2}+\beta_-^{IR}r_*^{-(1-\delta_2^*)/2},\\
&\delta_1^*=\sqrt{1+2(k+m^2+4Jp_{10}^2)/3},\\
&\delta_2^*=\sqrt{1+2(m^2-k+4Jp_{10}^2)/3}.\\
\end{split}
\end{equation}
Here $p_{10}=\sqrt{3}/(m^2+k)$,  $\alpha_{\pm}^{IR}$ and $\beta_{\pm}^{IR}$ are all finite constants. As  a result, to have the antiferromagnetic instability requires that the parameters satisfy following conditions,
\begin{equation}\label{cond1}
\left\{
\begin{split}
&1+2(k+m^2+4Jp_{10}^2)/3>0,\\
&1+2(m^2-k+4Jp_{10}^2)/3<0.
\end{split}
\right.,
\end{equation}
which gives following constraint,
\begin{equation}\label{cond3}
J_c^+(k,m^2)<J<J_c^-(k,m^2),~\text{and}~k>0.
\end{equation}
with $J_c^\pm(k,m^2)=-(m^2+k)^2(m^2+3/2\pm k)/12$.

\begin{figure}
\includegraphics[width=0.4\textwidth]{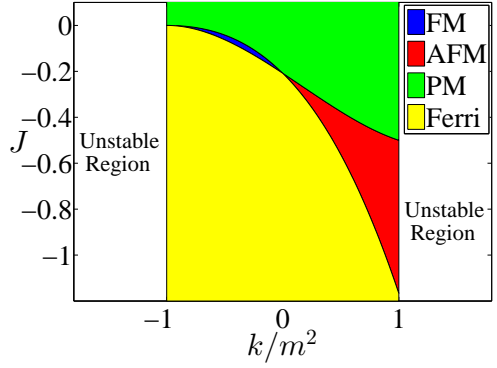}
\caption{Zero temperature phase diagram when $B=0$ and $m^2=1$. Here FM, AFM, PM and Ferri stand for  ferromagnetic phase, antiferromagnetic phase, paramagnetic phase and ferrimagnetic phase, respectively.}
\label{phase1}
\end{figure}

When external magnetic field $B=0$, for given mass square $m^2$, the zero temperature phase of the dual boundary then depends on the parameters $J$ and $k$. A typical example for $m^2=1$ is plotted in Fig.~\ref{phase1}. The red region corresponds to the case that parameters satisfy the conditions~\eqref{cond2} and \eqref{cond3}, which is what we will consider in this paper.

\subsection{N\'{e}el temperature and quantum criticality}

When the magnetic field is absent, i.e.,  $B=0$, under the conidtions~\eqref{cond0}, \eqref{cond2} and~\eqref{cond3},  the solution of $\alpha$ is always zero, which means there does not exist  spontaneous magnetization. But there is a critical temperature $T_{N0}$, lower than which a nonvanishing  $\beta$ begins to appear, which shows the system transits into antiferromagnetic phase and gives a nonzero spontaneous staggered magnetization.

On the other hand,  if turn on an  external magnetic field $B$, we have $\alpha\neq0$. When the external magnetic field is small, the value of $|\alpha|$ will increase with the increasing of $B$. Since the equation for $\beta$ couples with $\alpha$, the staggered magnetization is also influenced by magnetic field. This influence can be seen clearly from the effective mass square of $\beta$ in the IR region, namely, the near horizon region. Near the horizon, we have,
\begin{equation}\label{effm}
m^2_{\beta\text{eff}}|_{r_h}=\frac{4J\mu^2}{(k+m^2)^2}+m^2-k+\frac{64J^2\mu^2\alpha^2}{(k+m^2)^2(m^2-k)}.
\end{equation}
Under the restriction~\eqref{cond2}, we can see that, for small magnetic field, the effective mass square of $\beta$ in the IR region will be increased by magnetic field $B$, which leads the nonzero solution of $\beta$ to appear in lower and lower temperature with increasing the magnetic field. Thus the antiferromagnetic critical temperature $T_N$ will be suppressed by external magnetic field.

When the external magnetic is not very small, because of the  nonlinear coupling between $\beta$ and $\alpha$, it is not  easy to directly give the relation between effective mass square of $\beta$ and the external magnetic field. Instead, the behavior of $T_N$ with respect to magnetic field can be found numerically. Near the critical temperature, the staggered magnetization is very small, i.e., $\beta$ is a small quantity.  In that case we can neglect the nonlinear terms of $\beta$ in the equations~\eqref{eqab}, which leads to
\begin{equation}\label{eqab2}
\begin{split}
\alpha''+\frac{f'\alpha'}f-\frac{m^2_{\alpha\text{eff}}}{r^2f}\alpha=&\frac{B}{r^2f},\\
\beta''+\frac{f'\beta'}f-\frac{m^2_{\beta\text{eff}}}{r^2f}\beta=&0,
\end{split}
\end{equation}
with
\begin{equation}\label{eqabp2}
\begin{split}
&m^2_{\alpha\text{eff}}=4Jp_1^2+m^2+k,\\
&m^2_{\beta\text{eff}}=m^2-k+4Jp_1^2+8Jp_1\widetilde{p}_2\alpha^2,\\
&p_1=\frac{r^2\sqrt{3-B^2-4\pi T}}{(m^2+k)r^4-4J\alpha^2},\\
&\widetilde{p}_2=\frac{8J r^2\sqrt{3-B^2-4\pi T}}{(4J\alpha^2-m^2r^4)^2-k^2r^8}.
\end{split}
\end{equation}

When $T\neq0$, in the IR region, the regularity of the solution  gives following initial conditions
\begin{equation}\label{init1}
\left\{
\begin{split}
&\alpha'=\frac{\alpha m^2_{\alpha\text{eff}}+B}{4\pi T},\\
&\beta'=\frac{\beta m^2_{\beta\text{eff}}}{4\pi T},
\end{split}
\right.
\end{equation}
at the horizon. Without loss of the generality, we can set $\beta(r_h)=1$ due to the linearity of the equation of $\beta$.

When $T_N=0$, the initial conditions~\eqref{init1} imply that
\begin{equation}\label{Bc1}
B=-\alpha m^2_{\alpha\text{eff}}|_{r_h},\text{and}~m^2_{\beta\text{eff}}|_{r_h}=0,
\end{equation}
at the horizon. For given parameters $J, m^2$ and $k$, these two equations give the solution $\alpha(r_h)=\alpha_c$ and $B=B_c$. Here $\alpha_c$ is the initial value of $\alpha$ at the IR fixed point and $B_c$ is the critical magnetic field where the QPT occurs.

\begin{figure}
\includegraphics[width=0.52\textwidth]{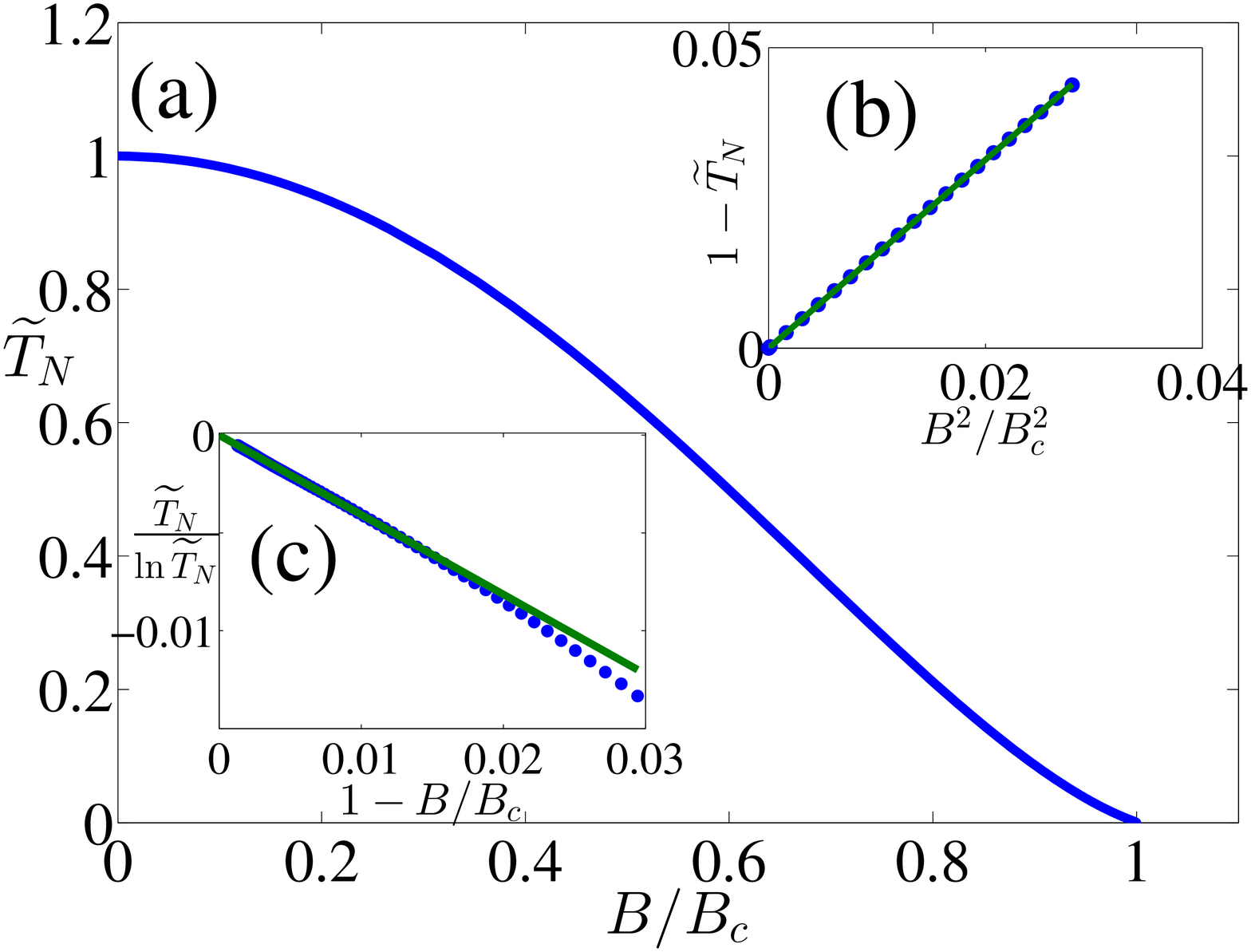}
\caption{The  antiferromagnetic critical temperature $T_N$ versus the magnetic field $B$. (a) In the whole region of $0\leq B\leq Bc$. (b) In the region of $B\ll B_c$. (c) In the region of $1-B/B_c\rightarrow0^+$.}
\label{TNB}
\end{figure}

In  Fig.~\ref{TNB}, we plot the N\'{e}el temperature $T_N$ in the case of $B<B_c$. Because different parameters satisfying restrictions \eqref{cond2} and~\eqref{cond3} give qualitatively same results, we here just show a typical example by taking $m^2=1, k=7/8$ and $J=-2/3$. With the increasing the magnetic field from zero to $B_c$, the N\'{e}el temperature decreases from $T_{N0}$ to zero. For small $B$, numerical results show that $T_N-T_{N0}\propto B^2$. When magnetic field approaches to $B_c$, we find that the N\'{e}el temperature is fitted well by following relation
\begin{equation}\label{TB1}
\widetilde{T}_N/\ln \widetilde{T}_N\simeq-0.4074(1-B/B_c).
\end{equation}
Here $\widetilde{T}_N=T_N/T_{N0}$ and $T_{N0}$ is the N\'{e}el temperature in the case of $B=0$.

The numerical results shown in~Fig.\ref{TNB} and the relation~\eqref{TB1} show that there is indeed a critical magnetic field $B_c$, at which the critical temperature for staggered magnetization, i.e. $\beta$, is zero. So a quantum phase transition occurs, which separates antiferromagnetic phase and a quantum disordered phase at zero temperature. We can also obtain other critical properties in the vicinity of QCP by numerical methods~\cite{Cai:2015a}. Especially, in the vicinity of QCP, this model gives a  dispersion relation for quasi-particle such as $\omega\sim q^2$ with energy $\omega$ and moment $q$, which shows that the boundary critical theory is a strong coupling theory with dynamic exponent $z=2$~\cite{Lohneysen.07}. This agrees with the results from condensed matter theory about antiferromagnetic metal~\cite{Lohneysen.07} and from the holographic model proposed by Ref.~\cite{Iqbal:2010eh}.

The behavior of the N\'{e}el temperature in the vicinity of QCP is quite non-trivial and needs to be further understood. The relation~\eqref{TB1} is not the usual power-law behavior in~\eqref{hyp2} or square-root form. However, it agrees with the predication in the 2-D QPTs in strong coupling case~\cite{Lohneysen.07,Hertz.76,Ar.Abanov}. The $d=z=2$ quantum critical theory is in general not in a weak coupling region for any $T>0$ and a strongly coupled effective classical model emerges that can be used to determine the critical dynamics~\cite{S.Sachdev2}.  As pointed out in Ref.~\cite{Cai:2015a}, it is just the reason that the  AdS/CFT correspondence is very suitable for this description. In addition, the original numerical results in Ref.~\cite{Cai:2015a} need to be confirmed, especial for  the logarithmic correction scaling rule Eq.~\eqref{TB1}. And we also need some more careful investigation to provide insight into physical properties in the vicinity of QCP in this holographic model. The most important is  to clarify the relationship between  this holographic model and the traditional theories of QPT and  to know what we can learn about real materials in experiment from our holographic results. These are our goals in the following sections.

\section{Emergent IR geometry and the behaviors near QCP}
\label{AFM2}
In order to understand the scaling relation~\eqref{TB1} and other numerical results presented in  Ref.~\cite{Cai:2015a}, we now pay attention to the background geometry in the zero temperature case. As the similar situation in Ref.~\cite{Faulkner:2009wj}, we will see that our numerical results about the behaviors in the vicinity of QCP are just controlled by the IR geometry which is governed by an emergent AdS$_2$ geometry.

\subsection{IR fixed point and critical magnetic field}
\begin{figure}
\includegraphics[width=0.5\textwidth]{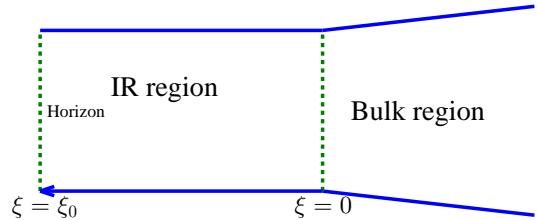}
\caption{A schematic figure for the inner IR region and outer bulk region. The horizon is located at $\xi=\xi_0$. The near horizon region is $\xi_0<\xi<0$, where an AdS$_2$ geometry emerges. The bulk region is not covered by $\xi$-coordinate. The solutions in the IR region and the bulk region should be matched at $\xi=0$.}
\label{FigIRUV}
\end{figure}

In the vicinity of QCP, the system is dominated by the features of IR region, where an AdS$_2$ geometry emerges.  Following  Ref.~\cite{Faulkner:2009wj}, we introduce a new length scale $\widetilde{r}$,
\begin{equation}\label{Qscal}
\mu^2+B^2=3\widetilde{r}^4
\end{equation}
Then the temperature of the black brane background can be written as,
\begin{equation}\label{Temp2}
T=\frac{3}{4\pi}(1-\widetilde{r}^4)
\end{equation}
Using this new scale, we define a new radial coordinate $\xi$ and time $\tau$ as,
\begin{equation}\label{scal1}
r-\widetilde{r}=\frac{\lambda}{6\xi},~~1-\widetilde{r}=\frac{\lambda}{6\xi_0},~~t=\lambda^{-1}\tau,~~~\lambda\rightarrow0.
\end{equation}
Use the variable $\xi$, we can define an inner IR boundary at $\xi\rightarrow\xi_0$ and an outer IR boundary at $\xi\rightarrow0^+$(see Fig.~\ref{FigIRUV}). The line element can be written as,
\begin{equation}\label{ds3}
ds^2=\frac{1}{6\xi^2}\left[-(1-\frac{\xi^2}{\xi_0^2})d\tau^2+(1-\frac{\xi^2}{\xi_0^2})^{-1}d\xi^2\right]+(dx^2+dy^2)
\end{equation}
Then the radial function $f(r)$ is,
\begin{equation}\label{fr}
f(\xi)=\frac{\lambda^2}{6\xi^2}(1-\frac{\xi^2}{\xi_0^2}).
\end{equation}
And the temperature with respect to $\tau$ is,
\begin{equation}\label{Temp3}
\widetilde{T}=\lambda^{-1}T=\frac1{2\pi\xi_0}.
\end{equation}
Here the case of zero temperature corresponds to an infinity $\xi_0$. Note that in the scaling limit~\eqref{scal1}, finite $\tau$ corresponds to the long time limit of the original time coordinate. Thus in the language of the boundary field theory,  the solution \eqref{ds3} corresponds  to the low frequency limit.

Since the radial coordinate can be considered as the resolution scale for the dual theory. The evolution of the geometry and the fields propagating therein along this radial direction represent the RG-flow of the dual field theory. Using the coordinate transformations~\eqref{scal1}, we can rewrite the equations~\eqref{eqab} in the limit of $\lambda\rightarrow0$ as
\begin{equation}\label{RGeq1}
\begin{split}
&\frac{d\alpha^2}{d\xi^2}-\frac{2\xi}{\xi_0^2-\xi^2}\frac{d\alpha}{d\xi}-\frac{\xi_0^2(B+\alpha m^2_{\alpha\text{eff}})}{6\xi^2(\xi_0^2-\xi^2)}=0,\\
&\frac{d\beta^2}{d\xi^2}-\frac{2\xi}{\xi_0^2-\xi^2}\frac{d\beta}{d\xi}-\frac{\xi_0^2m^2_{\beta\text{eff}}\beta}{6\xi^2(\xi_0^2-\xi^2)}=0.
\end{split}
\end{equation}
In the zero temperature case, which  corresponds to  $\xi_0\rightarrow\infty$, Eqs.~\eqref{RGeq1}  become
\begin{equation}\label{RGeq2}
\begin{split}
&\frac{d\alpha^2}{d\xi^2}-\frac{B+m^2_{\alpha\text{eff}}\alpha}{6\xi^2}=0,\\
&\frac{d\beta^2}{d\xi^2}-\frac{m^2_{\beta\text{eff}}}{6\xi^2}\beta=0.
\end{split}
\end{equation}
With the boundary conditions that $\alpha$ and $\beta$ are both finite at inner IR boundary and outer IR boundary or the existence of IR fixed points for $\alpha$ and $\beta$ in the IR region, these two equations have regular solutions only when $\alpha$ and $\beta$ are both constants. As a result,  we have,
\begin{equation}\label{RGeq2}
    B+m^2_{\alpha\text{eff}}\alpha=0,~~m^2_{\beta\text{eff}}=0.
\end{equation}
This  is just what we have obtained in Eqs.~\eqref{Bc1}. Note that in the case, $\alpha, \beta, p_1$ and $\widetilde{p}_2$ are all constants in IR region, so $\alpha m^2_{\alpha\text{eff}}$ and $m^2_{\beta\text{eff}}$ are also constants in IR region. With the restrictions that $m^2\geq|k|$ and $J$ satisfies Eqs.~\eqref{cond3}, we can conclude that when $B<B_c$, the solution with $\beta\neq0$ exists. But when $B>B_c$, there is only a trivial solution with $\beta=0$, by matching into bulk region, which will give a trivial solution such that $\beta(r)=0$ in the whole bulk region. Thus  there is a phase transition at $B=B_c$ at zero temperature, which divides an AFM phase from  a quantum disordered phase.

\subsection{Linear perturbations and dynamic exponent}
\label{app2}
In order to compute the magnetic fluctuations in the vicinity of QCP. We need to turn on the perturbations of two polarization fields. Because of the non-linear term in equations~\eqref{eomFM}, all the components couple with each other. As a result, we need consider the perturbations for all the components, i.e.,
\begin{equation}\label{pertbur1}
\begin{split}
&\delta M^{(a)}_{\mu\nu}=\epsilon C^{(a)}_{\mu\nu}e^{-i(\omega t+qx)},~~(\mu,\nu)\neq(r,y), (t,x)\\
&\delta M^{(a)}_{\mu\nu}=i\epsilon C^{(a)}_{\mu\nu}e^{-i(\omega t+qx)},~~(\mu,\nu)=(r,y), (t,x).
\end{split}
\end{equation}
Substituting  the ansatz~\eqref{pertbur1} into \eqref{eomFM} and compute up to the linear order of $\epsilon$, we can get the equations for perturbations. When $\omega, q\neq0$, however, we can't decouple all the equations.  In general, therefore we needs to solve the $6\times2$ components together in order to determine the dispersion relation. But if we consider the behavior for small frequency and wave vector in the vicinity of QCP, i.e., $\omega\rightarrow0$ and $q\rightarrow0$, the problem can be simplified. In that case, the equations for $C^{(a)}_{tx}$  and $C^{(a)}_{ry}$ decouple from the others and can be neglected.  Then we obtain $4\times2$ coupled equations,
\begin{equation}\label{pertur2}
\begin{split}
\left[(q^2+m^2)r^2-4JM^{(a)2}_{xy}\right]C^{(a)}_{tr}+q^2C^{(a)}_{rx}&\\
-8JM^{(a)}_{xy}M^{(a)}_{tr}C^{(a)}_{xy}+kr^2C^{(b)}_{tr}&=0,\\
\left(m^2-\frac{\omega^2}{r^2f}\right)C^{(a)}_{rx}+\frac{4JC^{(a)}_{ty}M^{(a)}_{xy}M^{(a)}_{tr}}{r^4f}+kC^{(b)}_{rx}&=0,\\
{C^{(a)}_{ty}}''-\frac{m^2C^{(a)}_{ty}}{r^2f}-\frac{kC^{(b)}_{ty}}{r^2f}+\frac{4JC^{(a)}_{rx}M^{(a)}_{xy}M^{(a)}_{tr}}{r^4f}&=0,\\
{C^{(a)}_{xy}}''+\frac{f'{C^{(a)}_{xy}}'}{f}+\left(\frac{\omega^2}{r^2f^2}-\frac{m^2+4JM^{(a)2}_{tr}}{r^2f}\right){C^{(a)}_{xy}}'&\\
-\frac{C^{(a)}q\omega}{r^4f^2}-\frac{kC^{(b)}_{xy}}{r^2f}-\frac{8JC^{(a)}_{tr}M^{(a)}_{xy}M^{(a)}_{tr}}{r^2f}&=0.
\end{split}
\end{equation}
with $(a, b)=(1, 2)$ or $(2, 1)$. In the paramagnetic phase, i.e., $M^{(1)}_{xy}=M^{(2)}_{xy}$ and $M^{(1)}_{tr}=M^{(2)}_{tr}$, the second and the third equations in~\eqref{pertur2} show that $C^{(1)}_{ty}=C^{(2)}_{ty}$ and $C^{(1)}_{rx}=C^{(2)}_{rx}$. Let $\widetilde{\beta}=(C^{(1)}_{xy}-C^{(2)}_{xy})/2$, then we obtain,
\begin{equation}\label{pereq3}
\widetilde{\beta}''+\frac{f'\widetilde{\beta}'}{f}+\left[\frac{\omega^2-Qq^2f}{r^2f^2} -\frac{m^2_{\beta\text{eff}}}{r^2f}\right]\widetilde{\beta}+\mathcal{O}(q^4)=0.
\end{equation}
with
\begin{equation}\label{eqQ1}
    Q=\frac{64J^2\alpha^2p_1^2}{[(m^2-k^2)r^4-4J\alpha^2]^2}>0. 
\end{equation}
As $q$ is infinitesimal, we can neglect the high order term $\mathcal{O}(q^4)$.  Note that no matter how small $\omega$ and $q$ are, we here can't neglect the $(\omega^2-Qq^2)/(r^2f^2)$  term, because it diverges when $f=0$.

In order to analyze the antiferromagnetic dispersion relation near the QCP for small frequency $\omega$ and wave vector $q$, we can take the IR-UV matching method proposed in Ref.~\cite{Faulkner:2009wj}. We first compute the retarded Green's function in the IR region. Use the metric~\eqref{ds3} in the limit of $\xi_0\rightarrow\infty$ and scale frequency as $\widetilde{\omega}=\lambda\omega$, we obtain,
\begin{equation}\label{IRdis1}
    \frac{d^2\widetilde{\beta}}{d\xi^2}+(\widetilde{\omega}^2-Q_0q^2/6\xi^2)\widetilde{\beta}=0,
\end{equation}
where we have used the results~\eqref{RGeq2}. $Q_0$ is the value of $Q$ in the IR region. Because $\alpha$ is a constant in IR region, $Q$ is also a constant in IR region. Solving this equation with the ingoing condition at $\xi\rightarrow\infty$, we have,
\begin{equation}\label{IRdis2}
\widetilde{\beta}\propto\sqrt{\xi}H^{(1)}_{\nu}(\widetilde{\omega}\xi).
\end{equation}
Here $H^{(1)}_{\nu}(x)$ is the  first Hankel function with order $\nu=\sqrt{9+6Q_0q^2}/6$. Using the properties of Hankel function and the fact that $q\ll1$, we can find that the asymptotic solution for $\beta$ in the outer IR region is,
\begin{equation}\label{IRdis3}
    \widetilde{\beta}\propto\xi^{-Q_0q^2/2}+\mathcal{G}(\widetilde{\omega})\xi^{1+Q_0q^2/2}
\end{equation}
with the IR retarded Green's function $\mathcal{G}(\widetilde{\omega})\propto\widetilde{\omega}^{Q_0q^2/3}$. Then using the method in Ref.~\cite{Faulkner:2009wj}, one can find that the retarded Green's function can be expanded in small $\omega$ as,
\begin{equation}\label{expand8}
\begin{split}
&G(\omega)\propto\\
&\frac{b^{(0)}_++\omega b^{(1)}_++O(\omega^2)+\mathcal{G}(\omega)(b^{(0)}_-+\omega b^{(1)}_-+O(\omega^2))} {a^{(0)}_++\omega a^{(1)}_++O(\omega^2)+\mathcal{G}(\omega)(a^{(0)}_-+\omega a^{(1)}_-+O(\omega^2))}
\end{split}
\end{equation}
with $b^{(i)}_\pm$ and $a^{(i)}_\pm$ are the analytical functions of $q^2$. By neglecting the irrelevant parts in~\eqref{expand8}, the retarded Green's function can be written as the  following form for small $\omega$ and $q$,
\begin{equation}\label{IRdis}
    G(\omega)=\frac{Z}{-q^2+a_1\omega+a_2\omega^{Q_0q^2/3}}
\end{equation}
with some constant $Z$. Considering the fact that when $q=0$, the retarded Green's function has a pole at $\omega=0$, we have  $a_2=0$. Then the dispersion relation is determined by the equation,
\begin{equation}\label{IRdis4}
    -Q_0q^2+a_1\omega=0,
\end{equation}
which tells us  $\omega\sim q^2$. Thus  we obtain the  dynamic exponent $z=2$.

One should note that, in general, the coefficients appearing in~\eqref{expand8} have scaling dimensions, and so dose for $a_1$ in~\eqref{IRdis4}. From the expression~\eqref{IRdis4}, we find that $a_1$ has  scaling dimension 1. Thus under the scaling transformation such that $r\rightarrow\lambda r$, $(t,x,y)\rightarrow\lambda^{-1}(t,x,y)$, which induces the transformations such that $\omega\rightarrow\lambda\omega, q\rightarrow\lambda q$ and $a_1\rightarrow\lambda a_1$, the equation~\eqref{IRdis4} is invariant. As  a result, the dynamic exponent $z=2$ is compatible with scaling symmetry.

\subsection{The scaling relation of $T_N$ with  $B$ near the QCP}
\label{app3}
Now let us  find the the scaling relation of $T_N$ with $B$ in the region of $B\rightarrow B_c^-$. The method we will take is  similar to the one as in the previous subsection. We first  compute the Green's function at finite temperature in the IR region, and then match it with the bulk solution.

In the IR region, near the fixed point $B=B_c+\delta B$ with $\delta B\rightarrow0^-$, we can neglect the non-linear term of $\beta$ and can get the equations for finite $\xi_0$,
\begin{equation}\label{scalrule1}
\begin{split}
    \frac{d\alpha^2}{d\xi^2}-\frac{2\xi}{\xi_0^2-\xi^2}\frac{d\alpha}{d\xi}=\frac{\xi_0^2[B+m^2_{\alpha\text{eff}}\alpha]}{6\xi^2(\xi_0^2-\xi^2)},\\
    \frac{d\beta^2}{d\xi^2}-\frac{2\xi}{\xi_0^2-\xi^2}\frac{d\beta}{d\xi}=\frac{\xi_0^2m^2_{\beta\text{eff}}\beta}{6\xi^2(\xi_0^2-\xi^2)}.
    \end{split}
\end{equation}
Under the boundary conditions that $\alpha$ and $\beta$ are both finite at $\xi=0$ and $\xi_0$, the solutions for Eqs.~\eqref{scalrule1} are constants. Then we have solution such that,
\begin{equation}\label{scalrule1b}
B_c+\delta B+m^2_{\alpha\text{eff}}\alpha=0,~~~ \beta=0,
\end{equation}
Using Eqs.~\eqref{Bc1}, up to the order of $\delta B$, we find that Eqs.~\eqref{scalrule1b} give solution $\alpha=\alpha_0\equiv\alpha_c+\delta\alpha$ with
\begin{equation}\label{soluta0}
\delta\alpha=-\left.\frac{\delta B}{(m^2_{\alpha\text{eff}}+\partial m^2_{\alpha\text{eff}}/\partial\alpha)}\right|_{\alpha=\alpha_c}.
\end{equation}
In order to compute the retarded Green's function, we consider the  deviations from the fixed point with  $\beta=0+\beta_1$ and $\alpha=\alpha_0+\alpha_1$. Note that $\delta B\rightarrow0^-$, up to the linear order of $\beta_1$ and $\alpha_1$, we have the equations of the fluctuations
\begin{equation}\label{scalrule2}
\begin{split}
    \frac{d\alpha_1^2}{d\xi^2}-\frac{2\xi}{\xi_0^2-\xi^2}\frac{d\alpha_1}{d\xi}=\frac{\xi_0^2(m^2_{\alpha\text{eff}}+\partial m^2_{\alpha\text{eff}}/\partial\alpha)\alpha_1}{6\xi^2(\xi_0^2-\xi^2)},\\
    \frac{d\beta_1^2}{d\xi^2}-\frac{2\xi}{\xi_0^2-\xi^2}\frac{d\beta_1}{d\xi}=\frac{\partial m^2_{\beta\text{eff}}}{\partial\alpha}\frac{\delta \alpha\beta_1}{6\xi^2(\xi_0^2-\xi^2)}.
    \end{split}
\end{equation}
Here effective mass terms and their derivative are taken the value at $\alpha=\alpha_0$.
%
The solutions for $\alpha_1$ and $\beta_1$ can be expressed as hypergeometric functions as,
\begin{equation}\label{scalrule3}
\begin{split}
\alpha_1&=\sum_{\pm}(\frac{\xi}{\xi_0})^{\frac{1\pm\delta_\alpha}2}C_{\pm}^{(\alpha)}{_2}F_1[\frac{3\pm\delta_\alpha}4, \frac{1\pm\delta_\alpha}4;1\pm\frac{\delta_\alpha}2;\frac{\xi}{\xi_0}],\\
\beta_1&=\sum_{\pm}(\frac{\xi}{\xi_0})^{\frac{1\pm\delta_\beta}2}C_{\pm}^{(\beta)}{_2}F_1[\frac{3\pm\delta_\beta}4, \frac{1\pm\delta_\beta}4;1\pm\frac{\delta_\beta}2;\frac{\xi}{\xi_0}]
\end{split}
\end{equation}
with $\delta_\alpha=\sqrt{1+2(m^2_{\alpha\text{eff}}+\partial m^2_{\alpha\text{eff}}/\partial\alpha)/3}$ and $\delta_\beta=\sqrt{1+\frac23(\partial m^2_{\beta\text{eff}}/\partial\alpha)\delta\alpha}$. These hypergeometric functions have logarithmic divergency when $\xi/\xi_0\rightarrow1$. So the coefficients $C_{\pm}^{(\alpha)}$ and $C_{\pm}^{(\beta)}$ should counteract these detergency. Then we can get the retarded Green's functions in the outer IR region near $\xi\rightarrow0$ as,
\begin{equation}\label{scalrule4}
\begin{split}
    \mathcal{G}_{\alpha\alpha}(\widetilde{T})=(2\pi\widetilde{T})^{\delta_\alpha}\frac{\Gamma(\frac34+\frac{\delta_\alpha}4)\Gamma(\frac14+\frac{\delta_\alpha}4)\Gamma(1-\frac{\delta_\alpha}2)} {\Gamma(\frac34-\frac{\delta_\alpha}4)\Gamma(\frac14-\frac{\delta_\alpha}4)\Gamma(1+\frac{\delta_\alpha}2)},\\ \mathcal{G}_{\beta\beta}(\widetilde{T})=(2\pi\widetilde{T})^{\delta_\beta}.
    \end{split}
\end{equation}
Now we can use the IR-UV matching method to express the UV Green's functions.  Similar to the expression~\eqref{expand8}, we can get the Green's functions for $\alpha_1$ and $\beta_1$ with some coefficients $a_\pm^{(n)}$ and $b_\pm^{(n)}$ which should be analytical functions of $\delta B$.   Note that two retarded Green's functions have poles at $\delta B=T=0$,
 we have,
\begin{equation}\label{scalrule5}
\begin{split}
    G_{\alpha\alpha}=\frac{Z_1}{\delta B+c_1^{(\alpha)}T+\mathcal{G}_{\alpha\alpha}(T)c_2^{(\alpha)}}, \\ G_{\beta\beta}=\frac{Z_2}{\delta B+c_1^{(\beta)}T+\mathcal{G}_{\beta\beta}(T)c_2^{(\beta)}}.
    \end{split}
\end{equation}
Here $\mathcal{G}_{\alpha\alpha}$ is irrelevant and can be neglected. One should note $c_i^{(\beta)}$ are also the functions of $c_1^{(\alpha)}$ because the equation for $\beta$ in the bulk depends on $\alpha$ but the equation for $\alpha$ in the bulk is independent on $\beta$. In the limit of $\delta B=0$, we will find $\delta_{\beta}=1$, so the $c_1^{(\beta)}$ term and $c_2^{(\beta)}$ term  degenerate. In that case, as the similar as  the case pointed out in Ref.~\cite{Faulkner:2009wj}, a logarithmic term appears. And the relevant terms in Green's functions read
\begin{equation}\label{scalrule6}
    G_{\alpha\alpha}=\frac{Z_1}{\delta B+c_1^{(\alpha)}T}, ~~G_{\beta\beta}=\frac{Z_2}{\delta B+c_3^{(\beta)}(c_1^{(\alpha)})T\ln T}
\end{equation}
with two constants $Z_1$ and $Z_2$. The poles for them are defined by,
\begin{equation}\label{scalrule7}
    \delta B+c_1^{(\alpha)}T=0,~~\delta B+c_3^{(\beta)}(c_1^{(\alpha)})T\ln T=0.
\end{equation}
Now note the fact that $\delta B\rightarrow-\delta B, c_1^{(\alpha)}\rightarrow-c_1^{(\alpha)}$ will lead $\alpha\rightarrow-\alpha$, which will not affect the  solution of $\beta$,  $c_3^{(\beta)}$ must be a function of $c_1^{(\alpha)2}$. Then in the case of $T\rightarrow0$ and $\delta B\rightarrow0$, there is a self-consistent ansatz such as $c_3^{(\beta)}=d_1c_1^{(\alpha)2}+O(c_1^{(\alpha)4})$ for small $c_1^{(\alpha)2}$ and finite constant $d_1$.\footnote{In general, the expansion of $c_3^{(\beta)}$ at small $c_1^{(\alpha)2}$ has a constant term such that $d_0$. This constant term must be zero, otherwise, the equations~\eqref{scalrule7} will not have solution if $\delta B\neq0$.} Then we can see  that  the relevant parts of the equations~\eqref{scalrule7} can be expressed as,
\begin{equation}\label{scalrule8}
    \delta B+c_1^{(\alpha)}T=0,~~\delta B+c_1^{(\alpha)2}d_1 T\ln T=0.
\end{equation}
Solving these equations, we find $\delta B\propto T/\ln T$. As a result  we have $b\equiv\delta B/B_c\propto T/\ln T$ and analytically show the relation (\ref{TB1}).

\subsection{Energy gap  and correlation length in the quantum disordered phase}

In  Ref.~\cite{Cai:2015a}, numerical results show that there is a gapped antiferromagnetic excitation in the quantum disordered phase. The energy gap and correlation length  still obey the power law form~\eqref{hyp3} and \eqref{hyp2} with respect to the tuning parameter $b$ in the vicinity of QPT. These results can also be understood  from the point of view of the emergent AdS$_2$ geometry.

In order to compute the energy gap of antiferromagnetic excitation, we should turn on a perturbation for $\beta$ with frequency $\omega$. In the region of quantum disordered phase with $T=0$ and $\delta B=B-B_c\rightarrow0^+$, using the results in the previous section, the background and perturbation equations for polarization fields read,
\begin{equation}\label{energ1}
\begin{split}
\alpha''+\frac{f'\alpha'}f-\frac{m^2_{\alpha\text{eff}}}{r^2f}\alpha-\frac{B}{r^2f}&=0,\\
\widetilde{\beta}''+\frac{f'\widetilde{\beta}'}{f}+\left[\frac{\omega^2}{r^2f^2}-\frac{m^2_{\beta\text{eff}}}{r^2f}\right]\widetilde{\beta}&=0
\end{split}
\end{equation}
and $\beta=0$. The real frequency $\omega$ giving the pole for Green's function $G_{\beta\beta}(B)$ leads to the gapped quasi-particle excitation. To compute this Green's function, we take the similar method as in the previous subsections. In the IR region, considering  the AdS$_2$ scaling limit at zero temperature, the equation~\eqref{energ1} can be written as,
\begin{subequations}\label{energ2}
\begin{align}
   &\frac{d^2\alpha}{d\xi^2}-\frac{B_c+\delta B+m^2_{\alpha\text{eff}}\alpha}{6\xi^2}=0,\label{energ2a}\\
   &\frac{d^2\widetilde{\beta}}{d\xi^2}+\left(\widetilde{\omega}^2-\frac{m^2_{\beta\text{eff}}}{6\xi^2}\right)\widetilde{\beta}=0.\label{energ2b}
     \end{align}
\end{subequations}

Up to the linear order of $\delta B$, the equation~\eqref{energ2a} gives the following regular solution with the IR fixed point,
\begin{equation}\label{energ3}
\alpha=\alpha_c+\delta\alpha.
\end{equation}
Here $\delta\alpha$ is defined as~\eqref{soluta0}. The equation for $\widetilde{\beta}$ can be written as
\begin{equation}\label{energ4}
\frac{d^2\widetilde{\beta}}{d\xi^2}+\left[\widetilde{\omega}^2-\left(\frac{\partial m^2_{\beta\text{eff}}}{\partial\alpha}\right)_{\alpha=\alpha_c}\frac{\delta\alpha}{6\xi^2}\right]\widetilde{\beta}=0.
\end{equation}
Comparing it with the equation~\eqref{IRdis1} in the subsection \ref{app2} and note the relationship in~\eqref{soluta0}, one can easily  find that the Green's function for $\widetilde{\beta}$ has the following form,
\begin{equation}\label{energ4}
    G_{\beta\beta}(\omega)=\frac{Z_3}{-\delta B+e_1\omega}
\end{equation}
for small frequency $\omega$ and $\delta B\rightarrow0^+$ with  two nonzero constants $e_1$ and $Z_3$. Thus we obtain  the relation between energy gap $\Delta$ and magnetic field $B$ as
\begin{equation}\label{energ5}
\Delta\propto B-B_c,
\end{equation}

As to correlation length, it can be computed in a similar way by setting $\omega=0$ but turning on  the perturbation for $\beta$ with nonzero moment $q$.  In that case, we still have  equation~\eqref{energ2a}  but, the equation~\eqref{energ2b}  is changed to be
\begin{equation}\label{energ6}
\frac{d^2\widetilde{\beta}}{d\xi^2}-\frac{Q_0q^2+m^2_{\beta\text{eff}}}{6\xi^2}=0.
\end{equation}
For small $q$ and $\delta B$, it gives the Green's function  as,
\begin{equation}\label{colle1}
    G_{\beta\beta}(\omega)=\frac{Z_4}{q^2+e_2\delta B}=\frac{Z_4}{q^2+1/\xi^2},
\end{equation}
where $Z_4$ is another constant and $\xi$ is called correlation length. We then easily obtain
\begin{equation}\label{colle2}
   \xi\propto(B-B_c)^{-1/2},
\end{equation}
which is as the same as equation~\eqref{hyp2} with $\nu=1/2$.

We see that results~\eqref{colle2} and~\eqref{energ6} obey the universal scaling relations~\eqref{hyp2} for quantum criticality with $z=2$, i.e., $\Delta\propto|B-B_c|^{z\nu}$. This can be viewed as the self-consistency check  for our model and computations.

\section{Conclusions and Discussions}
\label{AFM3}
\subsection{Summarize the holographic model }

We have  investigated the quantum phase transition (QPT) from antiferromagnetic phase (AF) to quantum disordered phase (QD) and the criticality in the vicinity of quantum critical pointc(QCP) in the antiferroamgentic materials induced by magnetic field in a holographic model proposed in Ref.~\cite{Cai:2014jta}. In the model, the N\'{e}el temperature $T_N$ is suppressed by magnetic field $B$. We proved that there is a critical magnetic  field $B_c$, at which $T_N=0$ and a QPT occurs. In order to analytically study the critical behavior in the vicinity of QCP, we employed  the IR-UV matching methods in Ref.~\cite{Faulkner:2009wj},  and the results show that the analytical method confirms the numerical results obtained in Ref.~\cite{Cai:2015a}.  In particular,  we found  that the dynamic exponent $z=2$, which means that the boundary critical theory is indeed a strong coupling theory with effective dimension $d_{\text{eff}}=d+z=4$. As a result, the hyperscaling law is violated and logarithmic corrections appear near the QCP. An interesting observation is that the critical behavior of the QCP is governed by the $AdS_2$ geometry emerging  in the IR region of the background geometry.

\subsection{Compare the results with RG-flow}

In this subsection, we will review the results from field theory and compare them with the results from the holographic model. Our discussion follows  Ref.~\cite{Lohneysen.07}, the details can be found there. A similar discussion as what we will do but mainly for  the case of $d=z=2$ can be found in Ref.~\cite{J.Bauer}.

\begin{figure}
\begin{center}
\includegraphics[width=0.35\textwidth]{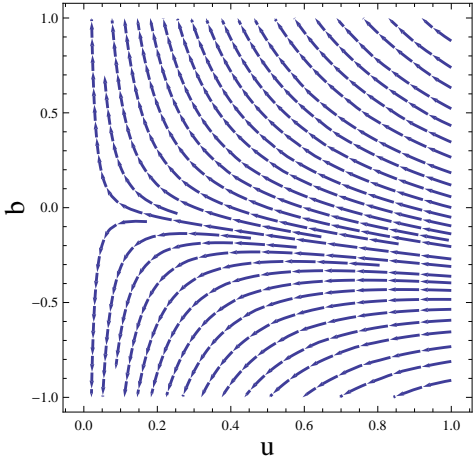}
\caption{The schematic RG-flow diagram of $u$-$b$ for $d+z\geq4$. There is a Gaussian fixed point at $u=b=0$.}
\label{Guass}
\end{center}
\end{figure}

In the original work by Hertz to quantum criticality~\cite{Hertz.76}, he considered the coupling between fermions and the low-energy bosonic field which condenses at the QCP. By integrating the fast fermions, the effective field theory is the Landau-Ginsburg-Wilson (LGW) $\phi^4$ theory with the upper critical dimension $d_c=4-z$, where $z$ is the dynamical exponent. Near the QCP, we can get the perturbative RG~\cite{A.J.Millis}. Let $u$, $T_N$, $\lambda$ and $b$ are the coupling constant, condensed temperature for bosonic field, scaling parameter and tuning parameter (in this paper, it's defined as $b=(B-B_c)/B$), respectively. Then the infinitesimal RG transformations are given by~\cite{A.J.Millis}
\begin{subequations}\label{RGM1}
\begin{align}
&\frac{d\mathcal{T}}{d\ln\lambda}=z\mathcal{T},~\label{RGM1a}\\
&\frac{db}{d\ln\lambda}=2b+uf_1(T,b),~\label{RGM1b}\\
&\frac{du}{d\ln\lambda}=(4-d-z)u-u^2f_2(T,b).~\label{RGM1c}
\end{align}
\end{subequations}
with some initial value $\mathcal{T}=T_N, b=b_0, u=u_0$. Here $f_1(\mathcal{T},b)$ and $f_2(\mathcal{T},b)$ are two functions of $\mathcal{T}, b$ defined by the details of model. At low $T$ close to the critical point $b=0$, $f_1(T,b)$ can be expressed as the power series of $T_N^2$ and $f_2(\mathcal{T},b)$ is constant. Thus we can set $f_1(\mathcal{T},b)=A_0+A_1T_N^2$ and $f_2(\mathcal{T},b)=f_2$. The RG equations~\eqref{RGM1} have a Gaussian fixed point at $\mathcal{T}=u=b=0$, which is unstable with respect to tuning parameter $b$, so a phase transition may occur at this point (see Fig.~\ref{Guass}). Since the RG equations are obtained by perturbations, we assume that the initial values satisfy $T_N, |b_0|, |u_0|\ll1$.
One of remarkable features is that this RG equation has a critical dimension at $d_c=d+z=4$. So, in general, when $d>4-z$, we can find that the boundary of ordered phase $T_N$ (such as the AF transition temperature, see Fig.~\ref{figTDg}), energy scale $\Delta$ in the quantum disordered phase and the correlation length $l$ in the vicinity of QCP obey the following scaling relations,
\begin{equation}\label{scaling1}
    T_N\sim (-b)^\psi,~~\Delta\sim b^{\nu z},~~l\sim b^{-\nu}.
\end{equation}
with $\psi=z/(d+z-2)$, $\nu=1/2$. However, this simple power law will lose its efficacy in the case of $d=z=2$, where the RG equations~\eqref{RGM1} need special consideration. In fact, logarithmic correction on $T_N$  appears. In order to see this, we directly solve $\mathcal{T}, u$ from the RG equation~\eqref{RGM1a},
\begin{equation}\label{RGM2}
    \mathcal{T}(\lambda)=T_N\lambda^z,~u(\lambda)=\frac{u_0}{1+u_0A_2\ln\lambda}.
\end{equation}
Next, integrating the scaling Eq.~\eqref{RGM1b} for $b(\lambda)$ and using Eqs.~\eqref{RGM1a} and~\eqref{RGM2}, we find
\begin{equation}\label{RGM3}
  b(\lambda)=\lambda^2\left[b_0+2e^{\frac2{u_0f_2}}E_1(2\ln\lambda+\frac2{u_0f_2})/(u_0f_2)\right].
\end{equation}
Here $E_1(x)$ is exponential integral function. For large real $x$, $E_1(x)\sim e^{-x}/x$. In the quantum critical region, using the scaling up to $\lambda\sim T_N^{-1/2}$, where $\mathcal{T}\sim1$, we have,
\begin{equation}\label{RGM4}
b(\lambda)=\frac{b_0}{T_N}+\frac1{f_2u_0\ln(1/T_N)}+O(\frac1{T\ln T}).
\end{equation}
Thus we see that the scaling relation depends on the value of $b(\lambda)$ as $\lambda\rightarrow\infty$. However, the concrete  form of $b(\lambda)$ can not be found from the perturbational RG equations directly. Because  the feature of strong coupling  in the case of $z=d=2$, the perturbational methods can't give complete information of the system. We see that   the perturbation method fails in this case to give a complete result. If $b(\lambda)$ is a finite constant, then we can find a linear relation $b_0\sim T_N$, which is  just the usual power law. If $b(\lambda)\ln\lambda\rightarrow0$ as $\lambda=1/\sqrt{T_N}\rightarrow\infty$, then we can get $b_0\sim T_N/\ln T_N$. Comparing with the usual power law, there is an additional logarithmic correction, just as the same as what we have obtained in the holographic model.

We see that, with some additional assumption, the perturbational RG equations give the same results as the holographic model. In fact, in the case of $d=z=2$, the low-energy  Fermi liquid leads to long-range order-parameter interaction. As a result, the coefficients of the high-order interactions in the LGW functional diverge, leading to an infinite number marginal operators~\cite{Ar.Abanov}. We see that the perturbational field method in this case fails to give out a complete description. However, the holographic setup can give self-consistent and complete predictions.


\subsection{Apply to real materials}
Now let's turn our attention into real materials and  see what we can learn from this holographic model about the real materials.
Here we focus on Er$_2$Ti$_2$O$_7$, as a kind of Magnetic pyrochlore oxides materials, which has  amazing properties.  At low temperatures  there exists an antiferromagnetic order arising with a N\'{e}el temperature of 1.2 K. The compound is complex and there together with the  nearest neighbor exchange,  the Dzyaloshinskii-Moriya, dipolar and magnetoelastic interactions do exist and play an important role in the formation of the low temperature antiferromagnetic state. There is spins ordered on the vertex-sharing tetrahedra, in non-coplanar structure. There order is formed  by quantum disorder, i.e., in a very unconventional way~\cite{Niven}. To describe this material the XY-like theoretical model where local spins are coupled by near-neighbour antiferromagnetic exchange and long-range dipole interactions on the pyrochlore lattice has been proposed~\cite{Zhitomirsky}. The key role in the model is played by the isotropic exchange between transverse components of near-neighbour spins and bond-dependent exchange anisotropy.  The authors have shown that the interplay of these interactions dictates the formation of the ground state by the quantum disorder mechanism. The inclusion of the exchange anisotropy reproduces well the observed second-order magnetic phase transition in Er$_2$Ti$_2$O$_7$ at zero field.  However, in Ref.~\cite{Oitmaa} it was argued that the dipolar interactions makes an important  contribution into the formation of the second order phase transition observed~\cite{Niven}. Probably  as argued in \cite{Zhitomirsky}  the exchange anisotropy may overcome the dipolar interaction at zero magnetic field where the tetrahedral magnetic units have no net magnetic moment in the ground state. However with the field when the canted states are formed the dipolar-dipolar interaction, which certainly exists between large magnetic moments associated with free Er$^{3+}$ ions ($J=15/2$, $g_J=6/5$), may become very important and can not be neglected. Therefore, the field evolution of the antiferromagnetic state and the role of the long-range part of the dipolar interactions in such an evolution remained unclear and beyond the scope of the discussed models. Here we proposed a holographic model which effectively absorbs all existing interactions in Er$_2$Ti$_2$O$_7$  in a form of strongly interacting tensor fields embedded in AdS space of a higher dimension. The extra dimension play a role of the renormalization parameter,  which  separates relevant and irrelevant degrees of freedom in the system. The proposed holographic model describes well not only the zero temperature thermodynamic phase transition and spin wave excitation spectrum  in Er$_2$Ti$_2$O$_7$ but also their evolution in the applied magnetic field~\cite{Cai:2015a}. The model may uncover an intriguing picture of many possible coexisting phases, which can arise at such complexity of  interactions  existing in Er$_2$Ti$_2$O$_7$~\cite{Ruff}.

In particular, in Ref.~\cite{Ruff} it was  found  that with magnetic order there arise well defined spin wave excitations. With applied magnetic field their spectrum is softening (see the spin wave spectrum measured at the field 1.5 T on the Fig.4 in  Ref.~\cite{Ruff}).  At the studies of an array of magnetic particles where the dipolar interaction is usually relevant it was shown that its long range character may lead to a formation of a landscape of low-energy states~\cite{Forrester,Kurten}. With increasing field the spins form canted state and the  energy landscape is flattering and there may arise  continuous switching between these states.  Therefore  such evolution may lead to a quantum critical behavior  and an appearance of quantum critical point QCP. Indeed the analysis of the low temperature phase diagram of Er$_2$Ti$_2$O$_7$  made in  Ref.~\cite{Ruff} supports an existence of the  rich variety of elementary excitations as the ground state is tuned by external magnetic field. The authors of  Ref.~\cite{Ruff} also suggested that most dramatic variations of the spectra with field arising at 1.5 T  may possibly correspond to the formation of a  continuous QCP.  Indeed in the framework of our holographic model similar  picture  emerges. The holographic model predicts that the magnetic order continuously changes with the field and  the QCP arises  at critical field $B_c\simeq1.45$ T.
\begin{figure}
\includegraphics[width=0.35\textwidth]{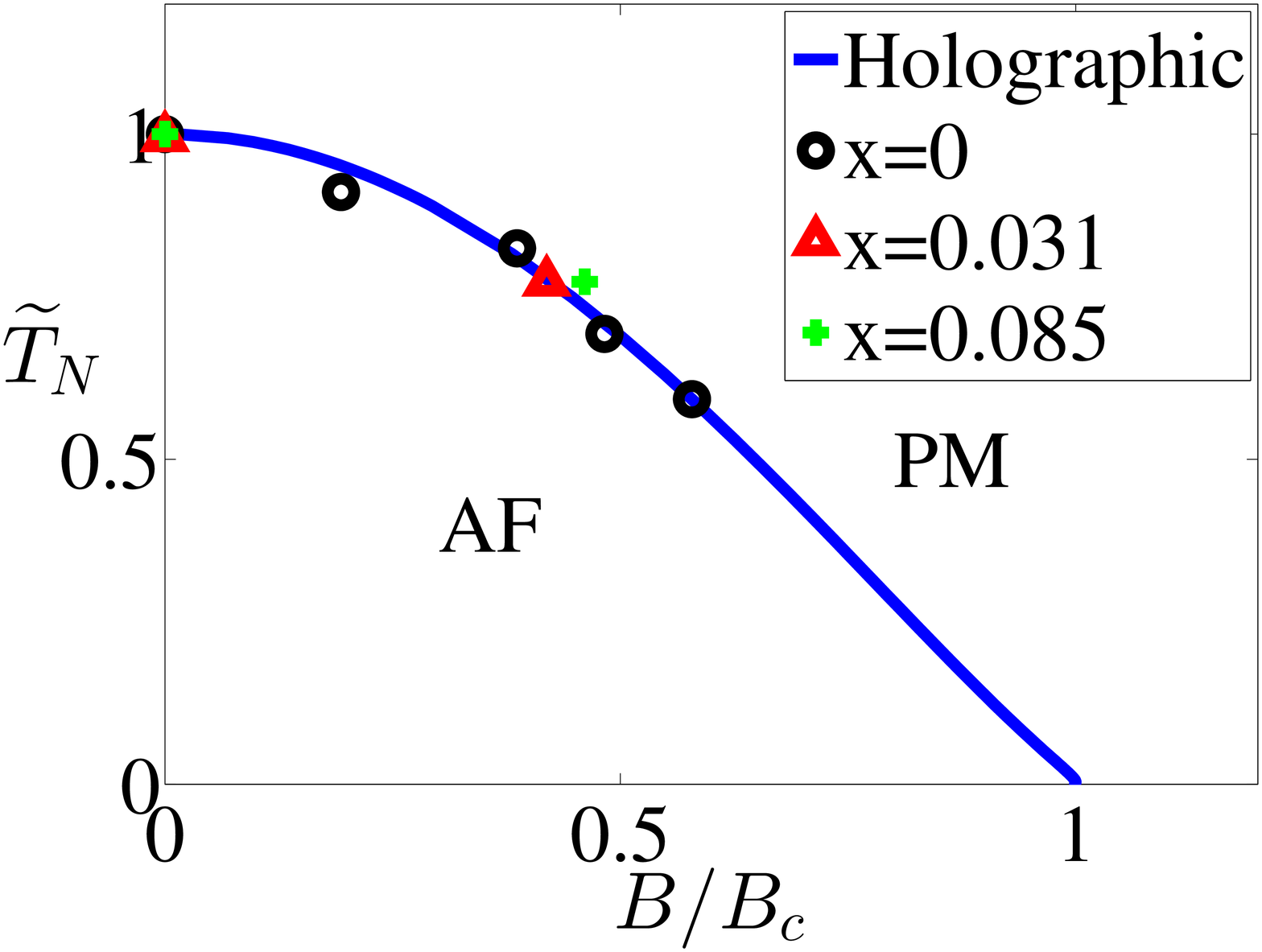}
\caption{The antiferromagnetic critical temperature $T_N$ versus the external magnetic field $B$ calculated at the best fitting model parameters $m^2=1, k=7/8$ and $J=-0.71$ (solid curve) and compared with
experimental data for pyrochlore compounds:  Er$_{2-2x}$Y$_{2x}$Ti$_2$O$_7$. Black circles corresponds to zero doping, $x=0$;
 magenta triangulars  - to $x=0.031$; green crosses - to $x=0.085$. The experimental data are from~\cite{Niven} and the figure is from~\cite{Cai:2015a}.}
\label{TNB3}
\end{figure}
Our model provides not only qualitative but also quantitative description of  the experimental data~\cite{E.Mathews,Niven} of the N\'{e}el temperature lowering with magnetic field (see, Fig.~\ref{TNB3}) and predict the QCP at 1.45T, which is very closed to the experimental results. There magnetic spins become partially aligned in the direction of the magnetic field.  Therefore the system requires less thermal energy to destroy the remaining magnetic spins order. We  found that similar behavior exist in other pyrochlore compounds which may be obtained from Er$_2$Ti$_2$O$_7$ by Y$^{3+}$ doping. Both doping and the increasing of magnetic field decrease the N\'{e}el temperature of the Y$^{3+}$-doped and undoped  Er$_2$Ti$_2$O$_7$ very similar until the QCP arises (see,  Fig.~\ref{TNB3}).  We see from  this figure that the critical temperature versus critical field curve is independent of the diamagnetic doping at the magnetic site.  This fact that the influence of dilution in (Er$_{1-x}$Y$_x$)$_2$Ti$_2$O$_7$ solid solutions is similar to the increase of magnetic field for pure Er$_2$Ti$_2$O$_7$  one more time stresses the importance and universality of complex magnetic interactions in these systems which can be well described in the framework of the proposed holographic model.

Here we also found that the high-field ground state does not correspond to a simple disordered paramagnet. The high-field state is characterized  by the intense dispersive excitation, which dispersion increases with the field. This was exactly observed in  Ref.~\cite{Ruff} where  it was found that the critical fluctuations of the low-field state are accompanied by a precursor of the intense dispersive excitation noticed at high field.
\begin{figure}
\includegraphics[width=0.4\textwidth]{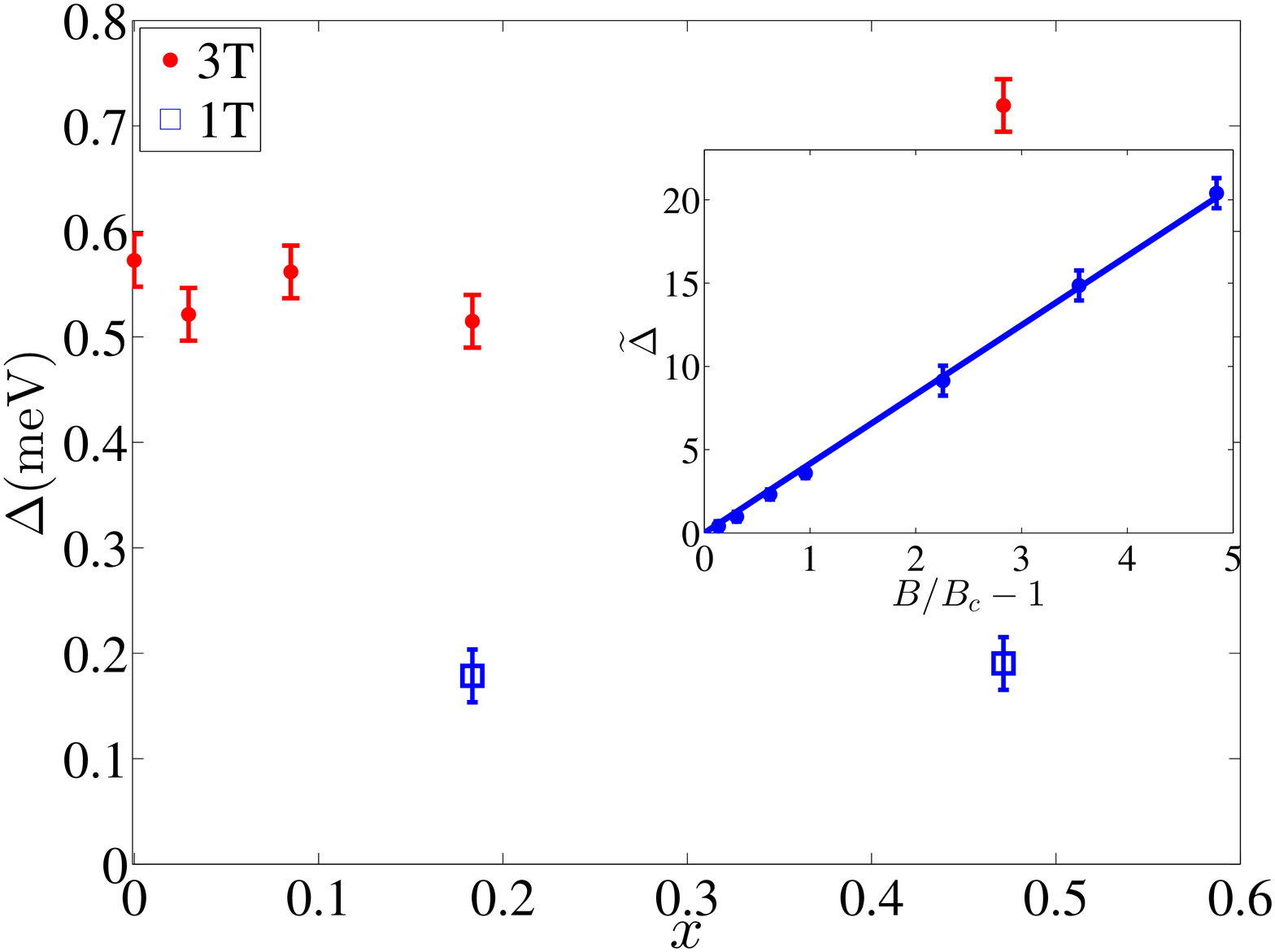}
\caption{The gap energy,$\Delta$, of Er$_{2-2x}$Y$_{2x}$Ti$_2$O$_7$ as a function of magnetic dilution ($x$) and magnetic field. The data is from Ref.~\cite{Niven}.}
\label{gapdata}
\end{figure}
In Fig.~\ref{gapdata}, we show the magnetic gap energy of Er$_{2-2x}$Y$_{2x}$Ti$_2$O$_7$ as a function of magnetic dilution ($x$) and magnetic field~\cite{Niven}. One should note that, in  Fig.~4 of Ref.~\cite{Ruff}, the excitation gap energy is zero when $B<B_c$ and $T\rightarrow0$. But by the magnetic heat capacity method in  Ref.~\cite{Niven}, there is a nonzero gap energy ($\simeq0.18$meV) when $B<B_c$ (see the case of $B=1$T in  Fig.~\ref{gapdata}). One can see that this nonzero gap energy is nearly independent on the doping $x$ which influences the antiferromagnetic electronic properties. So this nonzero gap in fact is from the background magnetic heat capacity which has nothing to do with the gap energy of the antiferromagnetic exaction. After removing this background contribution, we can find that the energy gap just satisfies the linear relationship~\eqref{energ5},  which is equivalent to
\begin{equation}\label{Bgap1}
  \widetilde{\Delta}\propto B/B_c-1
\end{equation}
with $\widetilde{\Delta}=\Delta/T_N$. In the subfigure of Fig.~\ref{gapdata} we plot the experimental data and the fitting curve by Eq.~\eqref{Bgap1}. The slope of the fitting curve is 4.2. We can also obtain the this slope from the holographic approach under the parameters of the best fitting from the Fig.~\ref{TNB3}, which gives that the slope is 5.0. So we see that our holographic  model can not only fit well the $T_N$-$B$ behavior but also predict the critical magnetic field which is very close to the measurement  result and the linear behavior of the energy gap with magnetic field with the slope nearly to the measurements at the same time.

We  also found that holographic model describe well magnetic properties of other class of materials, which have strong magnetic anisotropy  and large magnetic moments. The most illustrative example is BiCoPO$_5$~\cite{E.Mathews,Cai:2015a}. The compound  consists of  the interacting spins chains  giving the 1D character to the compound.  The spins have large magnetic moments, about  5.2 $\mu_B$ per Co  ion  measured at 1.5 K ~\cite{E.Mathews}. It is much higher than the value 3.9 $\mu_B$  expected for Co$^{2+}$  ions having S=3/2. Such large value of $\mu_{\text{eff}}$ is rather typical in Co$^{2+}$ containing compounds due to additional orbital magnetic moment created  in oxygen octahedra existed around the Co$^{2+}$  ions.

Magnetic structure has been determined with the powder neutron diffraction experiments~\cite{Mentre}.  It was suggested there the ferromagnetic pairs of Co$^{2+}$ ions are antiferromagnetically coupled within double chains. Because of these large values of the ferromagnetic pairs of Co$^{2+}$ ions there exists a dipole-dipole interaction between the resulting magnetic moments, which is, probably, significantly, weaker than the exchange interactions. There is also strong  the spin-orbital interaction, which always  exists in heavy atoms  such as Bi.  The competition of  all these interactions is probably responsible for the deviation from collinear  AF ground state in BiCoPO$_5$~\cite{Mentre}. In this sense the situation is very similar to the case of pyrochlore materials discussed above in the Fig.~\ref{TNB3}. Therefore, as above the long range large interacting moments  here can be replaced by the interacting tensor fields. It was found that the magnetic anisotropy and arrangements of magnetic moments in BiCoPO$_5$ can be continuously tuned by  temperature and external magnetic field~\cite{H.Yamaguchi}.

However, neither field induced order-disorder nor spin-flop transition was found in susceptibility and specific  heat measurements under external field, although the spin-flop transition was seen in the magnetization dependence on magnetic field  measured at T=2 K~\cite{E.Mathews}. The N\'{e}el temperature, $T_N$  and AF order were found to decrease gradually with the increase in field.  But even at high fields no extra features other than the N\'{e}el transition was observed in specific heat~\cite{E.Mathews} until the AF order was completely suppressed.  Such behavior is consistent with the prediction of our holographic model suggesting  that the quite strong and continuous change in magnetic anisotropy with the field may be well described with the interacting tensor field embedded in the gravitational AdS space, that is with the aid of the AdS/CFT duality.

\section*{Acknowledgements}
This work was supported in part by the National Natural Science Foundation of China  with grants No.11035008, No.11375247 and No.11435006.


\end{document}